\documentstyle[11pt]{article}
\begin{document}

\title{Topological mass term in effective Brane-World scenario with torsion}
\author{Nelson R. F. Braga\footnote{\noindent e-mail: braga@if.ufrj.br}$\,\,$
 and Cristine N. Ferreira\footnote{\noindent e-mail: crisnfer@if.ufrj.br}\\
\\
Instituto de F\'{\i}sica, Universidade Federal do Rio de
Janeiro,\\ Caixa Postal 68528, 21945-910, Rio de Janeiro, RJ, Brazil }

\date{\today}

\newcommand{\zero}{\setcounter{equation}{0}}

\renewcommand{\thesection}{\arabic{section}}

\renewcommand{\theequation}{\thesection.\arabic{equation}}

\maketitle

\begin{abstract}
In this work we investigate a braneworld scenario where a Kalb-Ramond  field 
propagates, together with gravity, in a five dimensional AdS slice.
The rank-2 Kalb-Ramond field is associated with torsion.
We study the compactification of the Kalb-Ramond field to
four dimensional space-time without gauge fixing, focusing on the effects of 
torsion. 
On the brane the Kalb-Ramond field interacts with  $A_{\mu}$-gauge  
and  scalar matter fields. 
We analyze the propagators of this theory and as an application 
investigate the consistency of such a model in the presence of a cosmic
string configuration.
One interesting feature of this model is the presence of topological charge
coming from a topological mass term that couples the gauge field 
$A_{\mu}$ with Kalb-Ramond modes. 
\end{abstract}

\section{Introduction}
The hierarchy problem is  one of the most instigating challenges for  
theoretical physicists. The disparity between
the electroweak and gravitational energy scales represents one of the
obstacles in the search for a  unified description of the fundamental
interactions.
On the other hand, consistency of string theory, our main present candidate to 
describe all fundamental interactions, requires our world to have more 
than four dimensions. 
Originally this extra dimensions were supposed to be very small (order of Plank length).
However, it has been proposed recently that the solution to the hierarchy 
problem may come from considering some of these extra dimensions to be not so 
small.  
In the approach of \cite{ADD} our space has one or more flat extra dimensions 
while, according to the so called Randall-Sundrum (RS) model\cite{RS}, 
hierarchy would be explained by one large warped extra dimension. 
In this RS approach our 4D world is a D3-brane embedded in a 5 dimensional 
Anti-de Sitter (AdS) bulk. Standard model fields are confined to the 3-brane while
gravity propagates in the bulk. In this scenario the hierarchy is generated 
by an exponential function of the compactification radius, called warp factor.

The Randall-Sundrum  model is defined in a 5-dimensional AdS slice
characterized by a background metric that may be written as

\begin{equation}
\label{1.1}
ds^2 = e^{-2 \sigma ( \phi ) } \eta_{\mu \nu} dx^{\mu} dx^{\nu } + r_c^2 d\phi^2
\end{equation}

\noindent
where $x^{\mu}$ are coordinates of the four-dimensional Lorentzian
surfaces at constant $\phi\,$ and  $\sigma ( \phi ) = kr_c|\phi|$.
The compact coordinate ranges from $ - \pi \le \phi \le \ + \pi $ with the points
$ ( x , \phi )$ and $ ( x , - \phi ) $ identified. 
In this approach gravity propagates in the bulk while the standard model fields 
propagate on the D-3 brane defined at
$\phi = \pi$. The energy scales are related in such a way that TeV mass scales 
are produced on the brane from Plank masses through the warp factor
$e^{-2kr_c\pi}$ 

\begin{equation}
m \, =\, m_0 \,e^{-2kr_c\pi}\,\,.
\end{equation}

\noindent In our case we will deal with massless fields in the bulk but this warp factor 
will appear in the effective couplings in four dimensions.

It has been recently proposed that torsion can play a non trivial role
in the RS scenario \cite{MSS}. The motivation to this approach is the fact that
both gravity and torsion are aspects of geometry. Since gravity propagates in the bulk
the same should be expected for torsion.
In this model the authors considered source of torsion to be a 
Kalb-Ramond (KR) field and showed that the zero mode would be 
exponentially suppressed by the AdS metric warp 
factor assuming a very weak value on the brane.  This would cause the 
illusion of a torsion free Universe since the massive KR modes that would have a larger 
coupling are very heavy.
However these massive modes could show up at TeV scales.

Torsion corresponds to the anti-symmetric part of the Affine connection
and shows up when one generalizes Einsteins theory of gravity
in such a way that the geometry is not only characterized by the curvature\cite{Gasp}.
Torsion can be related to coupling to fermionic fields\cite{Fer}, 
alternatively it can be associated with a  scalar field gradient for bosons 
in scalar tensor theories\cite{Valdir1,Valdir2} or
to a  rank 2 antisymmetric Kalb-Ramond tensor field as in \cite{MSS},\cite{MSS2}.

The anti-symmetric  KR tensor field was first introduced
within a string  theory context \cite{KR} where it is associated with
massless modes. The presence of such a background in string theory 
has the very important implication of non commutativity 
of space-time\cite{SW}. 
The KR tensor also appear in supersymmetric field theories.
More generally, p-form gauge fields appear in many supergravity models \cite{Gates}.

In this work we will consider a model where torsion is represented by 
a KR field in a Randall-Sundrum like scenario but with gauge and 
matter fields on a boundary D3-brane. One of the new features of our approach is that
we do not fix the KR gauge previous to compactification. This makes it possible
to  calculate the propagators for all the KR field components. 
Also we will include on the brane an alternative  topological mass term 
coupling KR and Abelian vector gauge fields that preserves Lorentz invariance. 
Furthermore we will analyze a possible  topological defect solution.
In particular we will consider a cosmic string 
configuration\cite{Vilenkin,Vilenkin1,Witten,Cristine4} although not taking 
gravitational effects into account as we will just focus on the effect of torsion. 
One finds objects analogous to these topological 
defects in condensed matter systems (where there is no gravity) for example 
in flux tubes in type-II superconductors\cite{Davis4}, or vortex filaments in a
superfluid\cite{Davis3}. Interesting discussions of cosmic strings
in brane world scenario can be found in Refs:\cite{Davis5,Pogosian1,Podolsky}
It is important to remark that KR fields have already been introduced in the 
cosmic string context with dilaton gravity\cite{Caroline2001,Kaloper} and in 
global vortex with extra dimensions\cite{Caroline2003}. 
They are an important ingredient to described vortex configuration in condensed 
matter systems \cite{Vilenkin2,Shellard,Davis4,Davis3}.
Important features appear when the model is supersymmetric\cite{Cristine1},
\cite{Cristine2}, \cite{Nogue}.

In section {\bf 2} we present the compactification scheme for the 
KR field in an AdS slice. In section {\bf 3} we study a model with gauge and matter 
fields on the brane with two kinds of coupling with the KR fields and 
calculate the propagators. Finally in section {\bf 4} we study a cosmic string 
configuration in this framework.

\section{The Kalb-Ramond field in brane world space time}

Let us consider a KR field $ B_{\hat \mu \hat \nu}$ 
in the 5-dimensional bulk (\ref{1.1})
with free dynamics described by

\begin{equation}
\label{ACTION}
S = \int dx^5\sqrt{-g_5}\frac{1}{6} \left\{
H_{\hat \mu \hat \nu \hat \rho}H^{\hat \mu \hat \nu \hat \rho}  \right\}
\end{equation}

\noindent where we are representing by indices with hat 
$\hat\mu , \hat\nu , ...= 0, ...4$ the bulk coordinates $( x^\mu\,,x^4 \equiv \phi )\,$
with $\mu = 0,..,3\,$ and $g_5 $ is the determinant of metric (\ref{1.1})
and the rank-3 antisymmetric field strength, identified with torsion is

\begin{equation}
H_{[\hat \mu \hat \nu \hat \rho]} = \partial_{\hat \mu}B_{\hat \nu \hat \rho }
+ \partial_{\hat \nu}B_{\hat \rho \hat \mu }\,+\,
\partial_{\hat \rho }B_{\hat \mu \hat \nu }\,\equiv\,
\partial_{[\hat \mu}B_{\hat \nu \hat \rho ]}
\end{equation}

The action (\ref{ACTION}) is invariant under the gauge transformation

\begin{equation}
\delta B_{\hat \mu \hat \nu}(x, \phi) = \partial_{\hat \mu}\xi_{\hat \nu}(x,\phi) -
\partial_{\hat \nu} \xi_{\hat \mu}(x,\phi)\,\,.
\end{equation}

Now, we follow an approach similar to \cite{MSS} (see also \cite{Ricardo})
in order to  find an effective action for the KR field on the four dimensional brane. 
The Lagrangian can be decomposed as

\begin{equation}
H_{\hat \mu \hat \nu \hat \rho}H^{\hat \mu \hat \nu \hat \rho} =
H_{\mu \nu \rho}H^{\mu \nu \rho} +
3 H_{4 \nu \rho}H^{4 \nu \rho}\label{b}\,\,.
\end{equation}

\noindent
In our approach we will rewrite  $B_{4 \mu}\,$ as 

\begin{equation}
B_{4 \mu}(x,\phi) = \partial_4 C_{\mu}(x,\phi) - \partial_{\mu} C_4(x,\phi)
\end{equation}

\noindent and the gauge invariant action gets

\begin{eqnarray}
{\cal S}_H &=& \int d^4x \int d\phi\, r_c e^{2\sigma}\Big\{ \eta^{\mu \alpha} 
\eta^{\nu \beta }
\eta^{\lambda \gamma} \frac{1}{6}H_{\mu \nu \lambda} H_{\alpha \beta \gamma}
\nonumber\\ &-&
\frac{1}{2r_c^2}e^{-2\sigma}\eta^{\mu \alpha}  \eta^{\nu \beta}\Big[
B_{\mu \nu}\partial_{\phi}^2B_{\alpha \beta} - 2C_{\mu\nu}
\partial_{\phi}^2B_{\alpha \beta}
+ C_{\mu \nu}\partial^2_{\phi} C_{\alpha \beta}\Big] \Big\}
\end{eqnarray}

\noindent
where, $C^{\mu \nu} \equiv \partial^{\mu}C^{\nu} -\partial^{\nu}C^{\mu}$.
This action is invariant under

\begin{eqnarray}
\delta C_\mu (x, \phi) &=&  \partial_\mu \omega (x, \phi) + \xi_\mu (x, \phi) 
\nonumber\\
\delta B_{ \mu \nu}(x, \phi) &=& \partial_{\mu}\xi_{ \nu}(x,\phi) -
\partial_{\nu} \xi_{\mu}(x,\phi)\label{transf1}
\end{eqnarray}

We now use a Kaluza-Klein decomposition for
the 4-D components of the KR field 

\begin{equation}
B_{\mu \nu}(x, \phi) = \sum_{n=0}^{n=\infty}B_{ \mu \nu}^n(x) \frac{\chi^n(\phi)}
{\sqrt{r_c}}\label{expansion}
\end{equation}

\noindent
and for $C_{\mu} $ and the gauge parameters

\begin{equation}
C_{\mu}(x,\phi) = \sum_{n=0}^{\infty} C_{\mu}^n(x)
\frac{\chi^n(\phi)}{\sqrt{r_c}}
\end{equation}

\begin{equation}
\xi_{\mu} (x, \phi)= \sum_{n=0}^{\infty}\xi_{\mu}^n (x) \frac{\chi^n(\phi)}{\sqrt{r_c}}
\end{equation}

\begin{equation}
\omega(x, \phi)= \sum_{n= 0}^{\infty}\omega^n (x) \frac{\chi^n(\phi)}{\sqrt{r_c}}
\end{equation}

\noindent
the corresponding gauge transformations for these modes are

\begin{equation}
\begin{array}{ll}
\delta C_{\mu}^n(x)  = \partial_{\mu} \omega^n (x) + \xi_{\mu}^n(x) \\
\delta B_{\mu \nu}^n = \partial_{\mu}\xi_{\nu}^n(x) - \partial_{\nu} \xi_{\mu}^n(x)
\end{array}
\end{equation}

\noindent Note that $C_{\mu}^n$ are not just Abelian gauge fields. They correspond 
to what are normally called Stuckelberg fields.

We now follow the approach of \cite{MSS},\cite{MSS2} of introducing normal modes
satisfying 

\begin{equation}
\label{eqmode}
-\frac{1}{r_c^2}\frac{d^2\chi_n}{d\phi^2} = m^2_n \chi_n e^{2 \sigma}\label{bessel1}
\end{equation}

\noindent and the orthonormality condition

\begin{equation}
\int e^{2\sigma} \chi^m(\phi) \chi^n(\phi) d \phi = \delta^{m n}\,\,.
\end{equation}

\noindent Then, integrating the coordinate $\phi$ we get the  effective action

\begin{eqnarray}
{\cal S}_H &=&\int d^4x \sum_{n=0}^{\infty}\Big\{ \eta^{\mu \alpha} \eta^{\nu \beta }
\eta^{\lambda \gamma}\frac{1}{6} H_{\mu \nu \lambda}^n H^n_{\alpha \beta \gamma}
\nonumber\\ &+& 
\frac{1}{2} m^2_n \eta^{\mu \alpha}  \eta^{\nu \beta}\left[
B_{\mu \nu}^n B_{\alpha \beta}^n - 2C_{\mu\nu}^n B_{\alpha \beta}^n
- C_{\mu \nu}^n C_{\alpha \beta}^n\right] \Big\} \label{SH}\nonumber\\
\end{eqnarray}

\noindent
where $H_{\mu \nu \lambda}^n = \partial_{[\mu }B^n_{\nu \lambda]}$.  

Although the modes $ \chi^n $ do not show up in the effective action,
in the next section we will consider the coupling of the KR fields 
to the brane fields. So it is important to find the explicit solutions for these 
modes to get their boundary values. 
We can re-write equation (\ref{eqmode}) in terms of the variables: 
$z_n = \frac{m_n}{k} e^{\sigma(\phi)}\,$ as

\begin{equation}
\Big( z_n^2 \frac{d^2}{dz_n^2} +  z_n \frac{d}{dz_n} + z_n^2\,\Big)  \chi^n =0
\end{equation}

\noindent The solutions, considering $ e^{-k r_c \pi}<<1 $ are 
\begin{equation}
\label{chi0}
\chi^0  = \frac{M^{3/2}}{M_pe^{k r_c \pi}}\,\,,
\end{equation}

\noindent  for the zero mode and 

\begin{equation}
\chi^n (z_n)  = {2 M^{3/2} J_0(z_n)e^{kr_c\pi }\over M_P \,\pi x_n }
\,\,,
\end{equation}

\noindent for $\,n\,\ge\,1\,$, where $ x_n\,=\, z_n (\pi) \,$ and 
$ M^2_P = M^3 [1- e^{-2kr_c\pi} ]/k \,$ relates the four dimensional Planck scale 
$M_P = 2 \times 10^{18} GeV $ to the (fundamental) Planck scale $M$\cite{RS}
so that $M_P$ is of the same order of $M$. 
The values of these modes on the boundary will be important in the next section
when the KR fields will couple to Abelian and matter fields there. 
We will represent them in the compact form 

$$ \chi^n (z_n (\pi)) \,=\, \chi^n (x_n) \,\equiv \chi^n_\pi.$$

We can see from the solution (\ref{chi0}) that the (constant) zero-mode 
$\chi^0$ exhibits 
a suppression by a large exponential factor. 
This was interpreted in \cite{MSS} as a possible explanation for the absence of an
observable torsion in our space-time.
Another very interesting consequence, also discussed in \cite{MSS}, would be the 
possibility of observing effects 
of the massive Kaluza Klein modes through new resonances in TeV-scale accelerators.

\section{Effective four dimensional theory and propagator analysis \zero }

Now we will consider the coupling of the KR field with the gauge and scalar 
matter fields that live on the four dimensional brane. We consider two kinds of 
couplings between the KR and the electromagnetic fields
located on the brane at $\phi = \pi$.  One comes from a topological mass term.
The other comes from the covariant derivative of  the scalar field that involves both 
the dual KR and the electromagnetic fields. Our four dimensional action is
 
\begin{equation}
\label{A4D}
{\cal S}_{4d}\,=\,\int dx^4  \Big[ -\frac{1}{2}D_{\mu}\Phi D^{\mu}
\Phi^{*}-\frac{1}{4}
F_{\mu\nu}F^{\mu\nu}-  {\xi \over 3}\,\epsilon^{\mu\nu \alpha \beta} A_{\mu}
\sum_{n=0}^{\infty}\,
\chi_{\pi}^n \, H^n_{\nu \alpha \beta}(x)
-\frac{\lambda}{4}( \vert \Phi\vert^2-\eta^2)^2 \Big]\,,
\end{equation}

\noindent where we just consider flat space as we are only interested in the 
effects of torsion. The coupling $\xi$ has dimension of $(mass)^{1/2}$.
The covariant derivative  and the field strengths are given by

\begin{eqnarray}
\label{CD}
D_{\mu}&=& \partial_{\mu} + iqA_{\mu} + i g \sum_{n=0}^{\infty} 
\chi_{\pi}^n \,\tilde H_{\mu}^n\\
F_{\mu \nu} &=& \partial_{\mu} A_{\nu} - \partial_{\nu}A_{\mu}\\
\nonumber\\
H^n_{\nu \alpha \beta } &\equiv&
\partial_{[\nu}B^n_{\alpha \beta ]}
\end{eqnarray}

\noindent
where $g$ is a coupling constant of dimension $(mass)^{-3/2} $ and 
$\tilde H_{\mu}^n$ is the mode expansion of the usual 4-dimensional 
KR dual of the field-strength given by

\begin{equation}
\tilde H_{\mu}^n \,\equiv\, 
\frac{1}{6} \epsilon_{\mu \nu \alpha \beta} H^{n \,\,\nu \alpha \beta }
\end{equation}

The gauge transformations that leave action (\ref{A4D}) invariant 
are 

\begin{equation}
\begin{array}{llll}
\Phi(x) &\rightarrow &\Phi(x)e^{i\Lambda (x)}, &\\
A_\mu(x) &\rightarrow & A_\mu (x) - \partial_\mu \Lambda(x),
&  \\
B_{\mu \nu}^n (x) &\rightarrow & B_{\mu \nu}^n(x) +
\partial_{\mu}\xi_{\nu}^n (x) - \partial_{\nu} \xi_{\mu}^n (x), &
\end{array}
\end{equation}

The effective  action on the brane
will include also the compactified KR action (\ref{SH}).

\begin{equation}
S_{eff.}\,=\, S_{4d} \,+\,S_H
\end{equation}

It should be noted that in this article we are not considering the curvature effects 
because we are only interested in the effects of the presence of torsion.

Now let us compute the propagators for the
gauge-field excitations, in a particular scalar field configuration. 
We take the  scalar field  $ \Phi $ to acquire the non-vanishing
vacuum expectation value $ < \vert \Phi \vert > = \eta$ corresponding to the
minimum of the potential. We parametrize $ \Phi $  as
\begin{equation}
\label{escalar}
\Phi =[\,\Phi (x)^{\prime }+\eta \,]\,e^{i\Sigma (x)},
\end{equation}

\noindent
where $ \Phi ^{\prime } $ is the quantum fluctuation around the ground state $\eta$.
In order to compute the propagators  we have to fix the gauge so as to make the action
non-singular. This is accomplished by the adding gauge-fixing terms

\begin{eqnarray}
\label{GF1}
{\cal L}_{A_{\mu} } &=&\frac{1}{2 \alpha }(\partial_{\mu} A^{\mu} 
+ 2 \alpha q \eta^2 \Sigma
)^2; \\[0.3cm]
\label{GF2}
{\cal L}_{B^n_{\mu \nu },C^n_{\mu}} &=& - \frac{1}{2\beta_n}
( \partial_{\mu}B_n^{\mu \nu} 
- 2 m_n^2\beta_n C_n^{\nu})^2\,.
\end{eqnarray}

Note that we are choosing this terms in such a way that the crossed terms 
involving $A_\mu$ and $\Sigma $ and also those involving 
$ B_n^{\mu \nu}$ and $ C_\mu$ are removed. So the total action   
in the non perturbed form of configuration  (\ref{escalar})  becomes

\begin{equation}
\begin{array}{ll}
{\cal L}_K = &-\frac{1}{4} F_{\mu \nu} F^{\mu \nu} +
\frac{1}{6}\left( \delta^{n n'}+ g^2 \eta^2 \chi^n_{\pi} \chi^{n'}_{\pi} \right) 
 H^n_{\mu \nu \kappa}  
H^{n' \mu \nu \kappa } - \frac{1}{3}( \xi + q \,g \eta^2 ) 
\chi^n_{\pi} \epsilon^{\mu \nu \alpha \beta }A_{\mu} H_{\nu \alpha \beta}^n  \\
&\\
&- \frac{1}{2}  m_n^2 (B^n_{\mu \nu}B^{n \mu \nu}  + C^n_{\mu \nu}C^{n \mu \nu}) + 
\frac{1}{2 \alpha}(\partial_{\mu}A^{\mu} )^2
+ 2 \alpha ( q \eta^2 \Sigma)^2  \\
&\\
& - q^2 \eta^2 A_{\mu} A^{\mu} - 
\frac{1}{2\beta_n} (\partial_{\mu}B_n^{\mu\nu})^2
- 2 \beta_n  ( m_n^2 C_n^{\nu})^2 - 
\eta^2 \partial_{\mu} \Sigma \partial^{\mu} \Sigma, \label{kin}
\end{array}
\end{equation}

\noindent Note that $\Sigma$ and $C_\mu$ decouple from the other fields
as a consequence of the gauge fixing terms (\ref{GF1}) and (\ref{GF2}) used.

The action has the general form
\begin{equation}
\label{opo}
{\cal L} = \frac{1}{2}\sum_{\alpha \beta} {\cal A}^{\alpha} {\cal O}_{\alpha
\beta} {\cal A}^{\beta},
\end{equation}

\vspace{0.3cm}

\noindent
where ${\cal A} _\alpha =(\Sigma ,A_\mu , B^n_{\mu \nu}, C^n_{\mu})$ 
and ${\cal O}_{\alpha \beta }$ is the wave operator. Representing the action 
in this form we see that calculating the propagators is equivalent 
to inverting the operator ${\cal O}$. This happens because we are only 
considering bilinear terms in the fields.

In order to invert the operator ${\cal O}$ we need to use 
an extension of the spin-projection operator formalism presented 
in \cite{Rivers,Nitsch}. 
In the present case we have to add other new
operators coming from the KR terms. The two projector operators which act on
the tensor field are:

\begin{equation}
(P^1_b)_{\mu \nu, \rho \sigma} = \frac{1}{2} (\Theta_{\mu \rho}\Theta_{\nu
\sigma} - \Theta_{\mu \sigma}\Theta_{\nu \rho}), 
\end{equation}

\begin{equation}
(P^1_e)_{\mu \nu, \rho \sigma} \,=\, \frac{1}{2} (\Theta_{\mu
\rho}\Omega_{\nu \sigma} - \Theta_{\mu \sigma}\Omega_{\nu \rho} 
\,-\,\Theta_{\nu
\rho}\Omega_{\mu \sigma} + \Theta_{\nu \sigma}\Omega_{\mu \rho}),
\end{equation}

\noindent
where $\Theta_{\mu \nu}$ and $\Omega_{\mu \nu}$ are, respectively, the
transverse and longitudinal projection operators, given by:

\begin{equation}
\Theta _{\mu \nu }=\eta _{\mu \nu }-\Omega _{\mu \nu },
\end{equation}

\noindent
and

\begin{equation}
\Omega_{\mu \nu} = \frac{\partial_{\mu} \partial_{\nu}}{\Box }.
\end{equation}

\noindent
We will need also the antisymmetric operator 

\begin{equation}
S_{\mu \gamma k} = \epsilon_{\lambda \mu \gamma k} \partial^{\lambda}.
\end{equation}

By using these operators the Lagrangian can be written as

\begin{equation}
\begin{array}{lll}
{\cal L} &= & A^{\mu} \left[\frac{1}{2}(\Box + 2 q^2 \eta^2) \Theta_{\mu \nu} 
- \frac{1}{2}(\frac{1}{\alpha}\Box - 
2 q^2 \eta^2) \Omega_{\mu \nu}\right]A^{\nu} + A^{\mu} 
\left[ (  \xi  + q \,g \eta^2 )\chi^n_{\pi} S_{\mu \nu k} \right] B^{\nu k \,\,n} \\
& &\\
& &+ B^{\alpha \beta\,\, n} \left[ ( m_n^2 \delta^{n n'} + (\delta^{n n'} + 
g^2 \eta^2 \chi^n_{\pi} \chi^{n'}_{\pi} )\,\Box )
\left(P^1_b\right)_{\alpha \beta. \nu k} + 
( m_n^2 + \frac{1}{4 \beta_n} \Box ) 
\left(P_e^1\right)_{\alpha \beta. \nu k} \right] B^{\nu k\,\, n'} \\
& &\\
& & + \Sigma \left[2 \alpha \eta^4 q^2 - \eta^2 \Box \right] \Sigma - 
m_n^2 C^{\mu} \left[ (\Box + 2 m_n^4 \beta_n) \Theta_{\mu \nu} + 
2 m_n^4 \beta_n \Omega_{\mu \nu}\right]C^\nu
\end{array}
\end{equation}

In order to find the wave operator$'$s inverse, we will need also the
products of operators for all non-trivial combinations involving the
projectors. The relevant multiplication rules are shown in table I.

\begin{table}
\caption{Multiplicative table}
\begin{center}
\begin{tabular}{|c|c|c|c|c|c|}\hline
&$\Theta^{\alpha \nu}$  & $\Omega^{\alpha \nu}$ &$S^{\alpha \nu \lambda }$& 
$(P_b)^{\alpha \nu , \lambda \beta} $&
$(P_e)^{\alpha \nu ,\lambda \beta} $\\
\hline
$\Theta_{\mu \alpha} $& $\Theta_{\mu}^{\,\,\,\,\nu}$ & 0 & 
$S_{\mu}^{\,\,\,\, \nu \lambda } $& 0& 0\\
\hline
$\Omega_{\mu \alpha }$&0 & $\Omega_{\mu }^{\,\,\,\,\nu} $& 0 & 0&  0  \\ 
\hline
$ S_{\lambda \mu \alpha} $& $ S_{\lambda \mu}^{\,\,\,\,\,\, \nu}$ & 0 & 
$\Box\Theta_\mu^{\,\,\,\,\nu} 
 $ & 2$S_{\mu}^{\,\,\,\,\nu \beta}$ & 0\\
\hline
$(P_b)_{\beta \lambda \,,\, \alpha \mu} $
& 0& 0& 2$ S_{\beta\,\,\,\mu}^{\,\,\,\nu}  $ & $ \Theta_{\mu}^{\,\,\,\, \nu} 
$& 0 \\
\hline
$ (P_e)_{\beta \lambda \,,\, \alpha \mu} 
$ & 0 & 0 & 0 & 0  & $\Omega_{\mu}^{\,\,\,\, \nu} $ \\
\hline 
\end{tabular}
\end{center}
\end{table}

\noindent
The part of the wave operator ${\cal O}$ eq. (\ref{opo}) that corresponds to the gauge 
fields $A_\mu , B^n_{\mu\nu}$ can be split into 
four sectors, according to:

\vspace{.3 true cm}

\begin{equation}
\label{olinha}
{\cal O}^\prime = \left(
\begin{array}{ll}
{\cal O}_{AA} & {\cal O}_{AB} \\
{\cal O}_{BA} & {\cal O}_{BB}
\end{array}
\right),
\end{equation}

\noindent
with

\begin{equation}
\begin{array}{ll}
{\cal O}_{A_{\mu}A_{\nu}} =  a_1\Theta_{\mu \nu} + a_2\Omega_{\mu \nu}\\
{\cal O}_{A_{\mu}B^n_{\nu\kappa}}= a_3^n S_{\mu \nu \kappa}\\
{\cal O}_{B_{\mu\nu}^n A_{\kappa}}= - a_3^n S_{\mu \nu \kappa} \\
{\cal O}_{B_{\alpha\beta}^n B_{\nu\kappa}^{n'}} 
= a_4^{n n'} (P^1_b)_{\alpha \beta, \nu \kappa} 
+ a_5^{n n'}(P^1_e)_{\alpha \beta , \nu \kappa},
\end{array}
\end{equation}

\noindent
where $a_1 , ..., a_5$ are:

\begin{equation}
\begin{array}{ll}
a_1 = \frac{1}{2}(\Box + 2q^2 \eta^2) ,\\
a_2 = - \frac{1}{2}( \frac{1}{\alpha} \Box - 2q^2 \eta^2),\\
a^n_3 = \frac{1}{2} ( \xi  + q \,g \eta^2 )\chi^n_{\pi} ,\\
a_4^{n n'} =  m^2_n \delta^{n n'} + 
(\delta^{n n' } + g^2 \eta^2 \chi^n_{\pi} \chi^{n'}_{\pi} )\,\Box , \\
a_5^{n n'} = (m_n^2 + \frac{1}{4\beta_n}\Box )\delta^{n n'},
\end{array}
\end{equation}

After some algebraic calculation, we find the inverse of the operator ${\cal O}^\prime $
of eq. (\ref{olinha}).
\begin{equation}
{\cal O}^{\prime\,\,-1} = \left(\begin{array}{ll}
X& Y\\
Z& W
\end{array}\right)
\end{equation}
\noindent
where the quantities X, Y, Z and W are:

\begin{equation}
\begin{array}{ll}
X = ({\cal O}_{AA} - {\cal O}_{AB} {\cal O}_{BB}^{-1} {\cal O}_{BA})^{-1},\\
Z = - {\cal O}_{BB}^{-1}{\cal O}_{BA} X,\\
W = ({\cal O}_{BB} - {\cal O}_{BA}{\cal O}_{AA}^{-1}{\cal O}_{AB})^{-1},\\
Y = - {\cal O}_{AA}^{-1}{\cal O}_{AB} W,
\end{array}
\end{equation}

\noindent
 or, explicitly

\begin{equation}
<A_{\mu}A_{\nu}> = \frac{i}{a_1 - 2a_3^n (a_4^{n n' })^{-1} a_3^{n'} \Box } 
\Theta_{\mu \nu} + \frac{i}{a_2}\Omega_{\mu \nu}
\label{A}
\end{equation}

\begin{equation}
<A_{\mu}B^n_{\nu\kappa}> = -2a_3^n(a_4^{n' {n'}'})^{-1}\frac{i}{a_4^{n^{''}n'} 
 - 2a_3^{n^{''}} a_1^{-1}a_3^{n'} \Box  } 
\epsilon_{\,\,\,\mu \nu  }^{\lambda \,\,\,\,\,\,\alpha\,\,}\partial_{\lambda } 
\Theta_{\alpha \kappa}\label{AB}
\end{equation}

\begin{equation}
<B^n_{\alpha \beta} B^{n'}_{ \gamma \kappa}> 
= \frac{i}{a_4^{nn'}  - 2a_3^n a_1^{-1}a_3^{n'} \Box } 
(P_b^1)_{\alpha \beta \gamma \kappa} + \frac{i}{ a_5^{n n'}} 
(P_e^1)_{\alpha \beta \gamma \kappa}\label{B}
\end{equation}

As the other sectors of the Lagrangian are diagonal we can calculate the other 
propagators by simply inverting the corresponding matrix elements.
For the $\Sigma - \Sigma$ propagator we find

\begin{equation}
<\Sigma \Sigma> = \left[ 2 \alpha \eta^3 a_2 \right]^{-1}\label{sigma}
\end{equation}

\noindent

For the $C^n_{\mu}$ fields the the operator ${\cal O}$ can be written as

\begin{equation}
\label{OC}
{\cal O}_{C^{n}_\mu C^{n'}_\nu} \,=\,
a^{n n'}_6  \Theta_{\mu \nu} +  a^{n n'}_7 \Omega_{\mu \nu}
\end{equation}

\noindent where 

\begin{equation}
\begin{array}{ll}
a^{n n'}_6 = - m_n^2 (\Box + 2 m_n^4 \beta_n)\delta^{n n'}\\
a^{n n'}_7 = - 2 m_n^6 \beta_n \,\delta^{n n'}
\end{array}
\end{equation}

\noindent in this expressions there is no sum over the index $n$.

Inverting the operator (\ref{OC}) we find the propagator for the $C_{\mu}$ 
fields

\begin{equation}
<C^n_\mu C^{n'}_\nu > = \frac{i}{a^{n n'}_6} \Theta_{\mu \nu} 
+ \frac{i}{a^{n n'}_7}\Omega_{\mu \nu}
\label{C}
\end{equation}
 
Let us now look for the zeros of the propagators that correspond to the effective 
masses. For the $A_\mu$ field the poles of eq. (\ref{A}) come from 
the zeros of the operator 
\begin{equation}
\label{AA}
 a_1 - 2 a_3^n (a_4^{n n'})^{-1} a_3^{n'} \Box 
\end{equation}

\noindent while for the $B_{\mu\nu}^n$ fields the poles of eq. (\ref{B})come from 
the zeros of
\begin{equation}
a_4^{n n'} - 2 a_3^n a_1^{-1} a_3^{n'} \Box  \label{BB}
\end{equation}   

Considering only the mode $\,n = 0\,$ of the KR field  we find 
from  equation (\ref{AA}) the momentum space poles of the propagator of
the gauge field $A_\mu$. 
\begin{equation}
K^2 = \left[ 2 q^2 \eta^2 - {(\xi + q \,g \,\eta^2\, )^2 \chi_\pi^0 \chi_\pi^0\over
(1 + g^2 \eta^2 \, \chi_\pi^0\chi_\pi^0 )} \right]\,.
\label{K}
\end{equation}
 
Similarly, for the  propagator KR field 
we find from eq.(\ref{BB}) 

\begin{eqnarray}
K^2 &=& 0 \label{m11}\\
K^2 &=& \left[ 2 q^2 \eta^2 - {(\xi + q \,g \,\eta^2\, )^2 \chi_\pi^0 \chi_\pi^0\over
(1 + g^2 \eta^2 \, \chi_\pi^0\chi_\pi^0 )} \right]\label{m12}
\end{eqnarray}

\noindent where again $ \chi^0_\pi\,=\,{M^{3/2} \over M_P} e^{-kr_c \pi }\,.$
So we see that for the KR field there is a massless solution (\ref{m11}) that may give 
rise to long range effects and a massive solution (\ref{m12}) with the same mass of 
the $A_{\mu} $ gauge field. 
 
\section{ Cosmic string configuration with Maxwell-Kalb-Ramond term in 
Brane World effective theory \zero}

In this section we study a cosmic string in the present 
Randall-Sundrum scenario. The model we will consider is a static cosmic string 
configuration living on the boundary D-3 brane in the presence of the AdS bulk 
KR torsion background.  
The effective action is 

\begin{eqnarray}
{\cal S}_M\, =\, \int dx^4\Big[&-&\frac{1}{2}D_{\mu}\Phi D^{\mu}\Phi^{*}-\frac{1}{4}
F_{\mu\nu}F^{\mu\nu}+  \frac{1}{6} \sum_{n=0}^{\infty} 
H^n_{\mu \nu \lambda} H^{\mu\nu\lambda\,\,\,n}
-\frac{\xi }{3} \epsilon^{\mu\nu \alpha \beta} \, A_{\mu }  \sum_{n=0}^{\infty} 
\chi_{\pi}^n  \,H^n_{\nu \alpha \beta}
\nonumber\\
&-& \frac{1}{2} \sum_{n=0}^{\infty}
 m^2_n \eta^{\mu \alpha}  \eta^{\nu \beta}\left[
B_{\mu \nu}^n B_{\alpha \beta}^n - 2C_{\mu\nu}^n B_{\alpha \beta}^n
+ C_{\mu \nu}^n C_{\alpha \beta}^n\right] \}
-\frac{\lambda}{4}( \vert \Phi\vert^2-\eta^2)^2
\Big]
\end{eqnarray}

The cosmic string configuration for the scalar field \cite{Nielsen} can be 
parametrized as usual in cylindrical coordinates $(t,r,\theta , z)$, where $r \geq 0$
and $0\leq \theta < 2 \pi $

\begin{equation}
\Phi  =\varphi (r)e^{i\theta }.
\label{vortex1}
\end{equation}

\vspace{0.3cm}

\noindent 
The boundary conditions for the fields $\varphi $ are

\begin{equation}
\begin{array}{ll}
\varphi ( r ) = \eta & r \rightarrow \infty \\
\varphi ( r ) = 0 & r=0
\end{array}
\label{cvortex1}
\end{equation}

\vspace{0.3cm}

The configuration for the gauge fields will be fixed by the
condition that the energy must be finite. 
This implies that at spatial infinity
($r \rightarrow \infty\,$) we must have
 
\begin{equation}
\label{finiteen}
D_{\theta} \Phi = \frac{1}{r}\left(\frac{\partial \Phi}{\partial \theta}\right) 
+ \frac{1}{r}\left( i q A_{\theta} + i g \sum_{n=0}^{\infty} 
\chi_{\pi}^n \,\tilde H_{\theta}^n \right)  \Phi =0\\
\end{equation} 

\noindent we take the following ansatz for the $\theta-$ component of  the gauge fields

\begin{eqnarray}
A_{\theta} &=& \frac{ 1}{q}(P(r)-1)\nonumber\\
\sum_{n=0}^{\infty} 
\chi_{\pi}^n \,\tilde H_{\theta}^n &=& {H (r)  \over g r} \,.
\end{eqnarray}

\noindent The condition (\ref{finiteen}) will be satisfied if 

\begin{eqnarray}
P(r) &=& 0 \hspace{1 true cm}  (r\rightarrow \infty)\nonumber \\
P(r) &=& 1 \hspace{1 true cm}  (r=0) \\
\nonumber\\
H(r) &=& 0 \hspace{1 true cm}  (r\rightarrow \infty)\nonumber \\
H(r) &=& 0 \hspace{1 true cm}  (r=0)
\label{cvortex3}
\end{eqnarray}

Without the KR field this configuration would be the ordinary 
one for a non charged static cosmic string that does not allow electric 
charge\cite{Nielsen}.
Now let us analyze how the presence of the KR field changes this picture.
For this we need the temporal component of the $A_\mu$ and ${\tilde H}_{\mu}\,$ fields.
We take the ansatz that they depend only on $r$: $A_{t}\,=\,A_{t}(r)$, 
$\sum_{n=0}^{\infty} \chi_{\pi}^n \,\tilde H_{t}^n \,=\,\left( \sum_{n=0}^{\infty} 
\chi_{\pi}^n \,\tilde H_{t}^n \right) (r) $. 
The finite energy condition for this component in the static case reads

\begin{equation}
D_t \Phi = \left( i q A_{t} + i g \sum_{n=0}^{\infty} 
\chi_{\pi}^n \,\tilde H_{t}^n \right)  \Phi \,=\, 0
\end{equation} 

\noindent
these equation is satisfied if the fields have the following asymptotic behavior

\begin{eqnarray}
A_t &=& 0 \hspace{1 true cm}  (r\rightarrow \infty) \nonumber\\
A_t &=& a  \hspace{1 true cm} (r=0)\\
\nonumber\\
\sum_{n=0}^{\infty} 
\chi_{\pi}^n \,\tilde H_{t}^n   &=& 0 \hspace{1 true cm} (r\rightarrow \infty )
\nonumber\\
\sum_{n=0}^{\infty} 
\chi_{\pi}^n \,\tilde H_{t}^n &=& h \hspace{1 true cm} (r=0)
\label{cvortex5}
\end{eqnarray}

We will consider in this first approach
to such a model that the background configuration does not involve the fields $C^n_\mu$.
That means: our ansatz involves $C^n_\mu\,=\,0$. 

In order to simplify the form of the equations of motion, let us use the notation
of the electric and magnetic fields, generalized also to the KR fields. 
The electrical type field 
$E^i$ and magnetic type field $B^i$ are defined as usual as

\begin{equation}
E^i = F^{0i} \hspace{2 true cm} B^i = - \epsilon^{ijk}F_{jk}\label{not1}
\end{equation}

\noindent where $i,j,k = 1,2,3\,$. For the KR we define 
\begin{equation}
{\cal E}^{(n) i} = - \epsilon^{i j k} H^{(n)}_{0 j k} \hspace{2 true cm} {\cal B}^{(n)} =
\epsilon^{i j k} H^{(n)}_{i j k}\label{not2}
\end{equation}

\noindent
which give us: $\tilde H^{(n) \mu} = ({\cal B}^{(n)}, \vec {\cal E}^{(n)}) $.
 
The equations of motion for the gauge fields, in terms of (\ref{not1}) and (\ref{not2}) 
get

\begin{equation}
{\vec \nabla} \times \vec E(r) = 0\label{camp1}
\end{equation}

\begin{equation}
{\vec \nabla} \cdot \vec B(r) =0 \label{camp4}
\end{equation}

\begin{equation}
{\vec \nabla} \cdot \vec E = - 2 \xi  \sum_{n=0}^{\infty}
\chi^n_{\pi} {\cal B}^n(r)  + \rho(r)\label{camp2}
\end{equation}

\begin{equation}
{\vec \nabla } \times \vec B(r) = 
-2 \xi  \sum_{n=0}^{\infty}
\chi^n_{\pi}   \vec {\cal E}^n(r) + \vec j(r) \label{camp3}
\end{equation}

\noindent
and the topological KR fields.

\begin{equation}
{\vec \nabla } \cdot \vec {\cal E}^n(r) =0
\end{equation}

\begin{equation}
\label{PN}
{\vec \nabla } \times \vec {\cal E}^n(r)= \xi \,\chi^n_{\pi}\, \vec B +
3\, m_n^2 \,{\vec {\cal P}}^n +   \frac{g \chi^n_{\pi} }{q} \,\nabla \times \vec{j}(r)
\end{equation}

\begin{equation}
{\vec \nabla} {\cal B}^n(r) = \xi \chi^n_{\pi} \vec E +
3 m_n^2 \vec {\cal H}^n + \frac{g \chi^n_{\pi}}{q} \nabla \rho
\label{camp5}
\end{equation}

\noindent
where ${\vec {\cal P}}^n = B_{0i}^n $, $\vec{\cal H }^n = \epsilon_{ijk} 
B_{ij}^n$
and the current $j^{\mu} = (\rho, \vec{j})$

\begin{equation}
j_{\mu} = - \frac{i q}{2}( \Phi^* D_{\mu}\Phi - \Phi D_{\mu} \Phi^* )\label{j}
\end{equation}

The topological current, divergenceless off shell, is given by

\begin{equation}
K^{\mu\,(n)} = \partial_{\nu}(\tilde F^{\mu \nu} + \tilde B^{\mu \nu\,(n)}).
\end{equation}

However, due to the absence of magnetic monopoles and to the zero order 
the first term can
be thrown away and we get

\begin{equation}
K^{\mu\,(n)} = \partial_{\nu} \tilde B^{\mu \nu\,(n)},
\end{equation}

\begin{equation}
\partial_{\mu} K^{\mu\,(n)} = 0
\end{equation}

The corresponding topological charged is defined as

\begin{equation}
Q_T^{(n)}\equiv \int K_t^{(n)} d^3x = \int d^3x \,{\cal B}^{(n)}\,\,,\label{top}
\end{equation}

\noindent
where ${\cal B}^{(n)}$ stands for the magnetic-like field (a scalar)
associated to the 2-form potential.  
In order to study the topological charge effect associated with the  KR 
field  we can write the ${\cal B}^{(n)}$  field as 

\begin{equation}
{\cal B}^{(n)} = \vec \nabla \cdot {\vec {\cal P}}^{(n)}.
\end{equation}

\noindent 
Then using definition (\ref{top}) and the divergence theorem in eq. (\ref{camp2}),
we define 

\begin{equation}
Q_T \equiv 2\, \xi\, \sum_{n=0}^{\infty} \chi_{\pi}^n Q_T^{(n)}
= 2\, \xi\, \sum_{n=0}^{\infty} \chi_{\pi}^n \int_V d^3x
 \vec \nabla . {\vec {\cal P}}^{(n)} = 2\,\xi\, \sum_{n=0}^{\infty} \chi_{\pi}^n
\int_S {\vec {\cal P}}^{(n)}. \hat u 
\hspace{.2 true cm} da,\,=\,Q  
\label{top1}
\end{equation}

\noindent where $Q$ stands for the electric charge 

\begin{equation}
Q = \int d^3x \rho \,=\,q\, \int d^3x \,\left(   q A_t \,+\, 
g \sum_{n=0}^{\infty} \chi_{\pi}^n {\tilde H}^n_t \right)\, \varphi^2
\end{equation}

From eq. (\ref{top1})  we find the behavior of the field ${\cal P}$

\begin{equation}
{\cal P}(r) \equiv 2\,\xi\,\sum_{n=0}^{\infty} \chi_{\pi}^n {\cal P}^n_r
 = \frac{1}{2 \pi }\frac{Q}{r} \label{intele2}
\end{equation}

\noindent  However, equation (\ref{PN}) tell us that, since 
outside the cosmic string $ \vec {\cal E}^n$ , $\vec B $
and $\vec{j}$ vanish, all the massive modes have a vanishing 
$ {\vec {\cal P}}^n$ also. So that only the massless mode has long range effects.
That means: outside the cosmic string:

\begin{equation}
{\cal P}(r) \equiv 2\,\xi\, \chi_{\pi}^0 {\cal P}^0_r
 = \frac{1}{2 \pi }\frac{Q}{r} 
\end{equation}

This result shows that there is a field configuration
outside the cosmic string that may cause some non trivial effects.

\section{Conclusion}

In this work, we analyzed the KR field in a five dimensional brane 
world scenario where it is associated with torsion. 
The theory is compactified to four dimensions without gauge fixing. 
As a result of this process we have a massless and a tower of massive 
modes corresponding to four dimensional Kalb-Ramond fields 
and also Stuckelberg fields on the brane. 
After the compactification we consider an effective field theory on the brane
including the free Kalb-Ramond sector plus matter and U(1) gauge fields.
Motivated by the interest in studying the effect of this Kalb-Ramond torsion background
in a cosmic string configuration we included in the model two types of Kalb-Ramond 
interactions, one in the U(1)-group covariant derivative  and the other 
given by a topological mass term. 

We also analyzed the propagators of this theory and found the Kalb-Ramond 
mass contribution to the $A_{\mu}$-gauge field caused by the 
topological mass term considering the zero mode. 
Finally we showed that it is possible to construct a topological configuration 
like a static cosmic string on the brane whose formation involves only the zero mode.

\vskip 1cm

\noindent{\bf Acknowledgments}: The authors are partially supported by  CNPq 
(Brazilian Research Agency).

\end{document}